\documentclass[11pt]{article}

\usepackage{amsmath,amssymb}

\usepackage{amsthm}

\usepackage{ifthen}

\usepackage{ifpdf}

\ifpdf
\usepackage[pdftex]{graphicx}
\usepackage[pdftex]{hyperref}
\else
\usepackage[dvips]{graphicx}
\fi

\ifthenelse{\isundefined{\NoGenerateLetterSize}}
{

\ifpdf
\pdfpageheight\paperheight
\pdfpagewidth\paperwidth
\fi
}
{}

\newcommand{\Comment}[1]{}

\long\def\LongVersion#1\LongVersionEnd{#1}
\long\def\ShortVersion#1\ShortVersionEnd{}

\Comment{
\setlength{\textheight}{8.9in}
\setlength{\textwidth}{6.6in}
\setlength{\evensidemargin}{-0.2in}
\setlength{\oddsidemargin}{-0.2in}
\setlength{\headsep}{10pt}
\setlength{\topmargin}{-0.3in}
\setlength{\columnsep}{0.375in}
\setlength{\itemsep}{0pt}
\parskip=0.06in
}

\newtheorem{theorem}{Theorem}[section]
\newtheorem{lemma}[theorem]{Lemma}
\newtheorem{observation}[theorem]{Observation}
\newtheorem{corollary}[theorem]{Corollary}
\newtheorem{proposition}[theorem]{Proposition}

\theoremstyle{definition}
\newtheorem{property}[theorem]{Property}
\theoremstyle{plain}

\Comment{
\newtheorem{theorem}{Theorem}
\newtheorem{lemma}[theorem]{Lemma}
\newtheorem{observation}[theorem]{Observation}
\newtheorem{corollary}[theorem]{Corollary}
\newtheorem{proposition}[theorem]{Proposition}

\theoremstyle{definition}
\newtheorem{property}[theorem]{Property}
\theoremstyle{plain}
}

\Comment{
\newtheorem{theorem}{Theorem}[section]
\newtheorem{lemma}{Lemma}[section]
\newtheorem{corollary}{Corollary}[section]

\newtheorem{proposition}{Proposition}[section]

\theoremstyle{definition}

\theoremstyle{plain}
}

\newenvironment{AvoidOverfullParagraph}[0]
{\sloppy\ignorespaces}
{\par\fussy\ignorespacesafterend}

\newcommand{\Metric}[0]{\ensuremath{\mathcal{M}}}
\newcommand{\Distance}[0]{\ensuremath{\delta}}
\newcommand{\WorkFunction}[0]{\ensuremath{\mathit{w}}}
\newcommand{\Algorithm}[0]{\texttt{Alg}}
\newcommand{\Optimal}[0]{\texttt{Opt}}
\newcommand{\ConfigurationDistance}[0]{\ensuremath{\mathit{D}}}

\begin{document}

\title{On the Additive Constant of the  $k$-server Work Function Algorithm}

\author{
Yuval Emek\thanks{Tel Aviv University, Tel Aviv, 69978 Israel.
E-mail: \texttt{yuvale@eng.tau.ac.il}.
This work was partially done during this author's visit at LIAFA, CNRS and
University Paris Diderot, supported by Action COST 295 DYNAMO.}
\and
Pierre Fraigniaud\thanks{CNRS and University Paris Diderot, France.
Email:  \texttt{pierre.fraigniaud@liafa.jussieu.fr}.
Additional support from the ANR project ALADDIN, by the INRIA
project GANG, and by COST Action 295 DYNAMO.}
\and
Amos Korman\thanks{CNRS and University Paris Diderot, France.
Email:   \texttt{amos.korman@liafa.jussieu.fr}.
Additional support from the ANR project ALADDIN, by the INRIA
project GANG, and by COST Action 295 DYNAMO.}
\and
Adi Ros\'{e}n\thanks{CNRS and University of Paris 11, France.
Email:  \texttt{adiro@lri.fr}.
Research partially supported by ANR projects AlgoQP and ALADDIN.}
}

\date{}

\maketitle

\begin{abstract}
We consider the Work Function Algorithm for the $k$-server problem
\cite{CL91,KP95}.
We show that if the Work Function Algorithm is $c$-competitive, then it is
also {\em strictly} $(2c)$-competitive.
As a consequence of \cite{KP95} this also shows that the Work Function
Algorithm is strictly $(4k-2)$-competitive.
\end{abstract}

\section{Introduction}

A (deterministic) online algorithm  \Algorithm{} is said to be
\emph{$c$-competitive} if for all finite request sequences $\rho$, it holds
that $\Algorithm(\rho) \leq c\cdot OPT(\rho) +\beta$, where $\Algorithm(\rho)$
and $OPT(\rho)$ are the costs incurred by \Algorithm{} and the optimal
algorithm, respectively, on $\sigma$ and $\beta$ is a constant independent of
$\rho$.
When this condition holds for $\beta=0$, then \Algorithm{} is said to be
\emph{strictly $c$-competitive}.

The $k$-server problem is one of the most extensively studied online problems
(cf. \cite{BE}).
To date, the best known competitive ratio for the $k$-server problem on
general metric spaces is $2k-1$ \cite{KP95}, which is achieved by the Work
Function Algorithm \cite{CL91}.
A lower bound of $k$ for any metric space with at least \( k + 1 \) nodes is
also known \cite{MMS90}.
The question whether online algorithms are strictly competitive, and in
particular if there is a {\em strictly} competitive $k$-server algorithm, is
of interest for two reasons.
First, as a purely theoretical question.
Second, at times one attempts to build a competitive online algorithm by
repeatedly applying another online algorithm as a subroutine.
In that case, if the online algorithm applied as a subroutine is not strictly
competitive, the resulting online algorithm may not be competitive at all due
to the growth of the additive constant with the length of the request
sequence.

In this paper we show that there exists a strictly competitive $k$-server
algorithm for general metric spaces.
In fact, we show that if the Work Function Algorithm is $c$-competitive, then
it is also strictly $(2c)$-competitive.
As a consequence of \cite{KP95}, we thus also show that the Work Function
Algorithm is strictly $(4k-2)$-competitive.

\section{Preliminaries}
Let \( \Metric = (V, \Distance) \) be a metric space.
We consider instances of the \(k\)-server problem on \(\Metric\), and when
clear from the context, omit the mention of the metric space.
At any given time, each server resides in some node \( v \in V \).
A subset \( X \subseteq V \), \( |X| = k \), where the servers reside is
called a \emph{configuration}.
The \emph{distance} between two configurations \(X\) and \(Y\), denoted by
\(\ConfigurationDistance(X, Y)\), is defined as the weight of a minimum weight
matching between \(X\) and \(Y\).
In every \emph{round}, a new \emph{request} \( r \in V \) is presented and
should be \emph{served} by ensuring that a server resides on the request \(r\).
The servers can move from node to node, and the movement of a server from node
\(x\) to node \(y\) incurs a \emph{cost}  of \(\Distance(x,
y)\).

Fix some initial configuration \(A_0\) and some finite request sequence
\(\rho\).
The \emph{work function} \(\WorkFunction_{\rho}(X)\) of the configuration
\(X\) with respect to \(\rho\) is the optimal cost of serving \(\rho\)
starting in \(A_0\) and ending up in configuration \(X\).
The collection of work function values \( \WorkFunction_{\rho}(\cdot) = \{ (X,
\WorkFunction_{\rho}(X)) \mid X \subseteq V, |X| = k \} \) is referred to as
the \emph{work vector} of \(\rho\) (and initial configuration \(A_0\)).

A move of some server from node \(x\) to node \(y\) in round \(t\) is called
\emph{forced} if a request was presented at \(y\) in round \(t\).
(An empty move, in case that \( x = y \), is also considered to be forced.)
An algorithm for the \(k\)-server problem is said to be \emph{lazy} if
it only makes forced moves.
Given some configuration \(X\), an offline algorithm for the \(k\)-server
problem is said to be \emph{\(X\)-lazy} if in every round other than the last
round, it only makes forced moves, while in the last round, it makes a forced
move and it is also allowed to move servers to nodes in \(X\) from nodes not
in \(X\).
Since unforced moves can always be postponed, it follows that
\(\WorkFunction_{\rho}(X)\) can be realized by an \(X\)-lazy (offline)
algorithm for every choice of configuration \(X\).

Given an initial configuration \(A_0\) and a request sequence \(\rho\), we
denote the total cost paid by an online algorithm \Algorithm{} for serving
\(\rho\) (in an online fashion) when it starts in \(A_0\) by \(\Algorithm(A_0,
\rho)\).
The optimal cost for serving \(\rho\) starting in \(A_0\) is denoted by \(
\Optimal(A_0, \rho) = \min_{X}\{ \WorkFunction_{\rho}(X) \} \).
The optimal cost for serving \(\rho\) starting in \(A_0\) and ending in
configuration \(X\) is denoted by \( \Optimal(A_0, \rho, X) =
\WorkFunction_{\rho}(X) \).
(This seemingly redundant notation is found useful hereafter.)
\Comment{
Note that the number of servers used by the algorithm (either online or
optimal) is implicitly cast in the above notation through the cardinality of
the initial configuration \(A_0\).
(This will be important when the number of servers is not explicitly stated.)
} 

Consider some metric space \(\Metric\).
In the context of the \(k\)-server problem, an algorithm \Algorithm{} is said
to be \emph{\(c\)-competitive} if for any initial configuration \(A_0\), and
any finite request sequence \(\rho\), \( \Algorithm(A_0, \rho) \leq c \cdot
\Optimal(A_0, \rho) + \beta \), where \(\beta\) may depend on the initial
configuration \(A_0\), but not on the request sequence \(\rho\).
\Algorithm{} is said to be \emph{strictly \(c\)-competitive} if it is
\(c\)-competitive with additive constant \( \beta = 0 \), that is, if for any
initial configuration \(A_0\) and any finite request sequence \(\rho\), \(
\Algorithm(A_0, \rho) \leq c \cdot \Optimal(A_0, \rho) \).
As common in other works, we  assume that the online algorithm and the optimal
algorithm have the same initial configuration.

\section{Strictly competitive analysis}
We prove the following theorem.

\begin{AvoidOverfullParagraph}
\begin{theorem} \label{theorem:OmitAdditiveTermWFA}
If the Work Function Algorithm is \(c\)-competitive, then it is also strictly
\( (2 c) \)-competitive.
\end{theorem}
\end{AvoidOverfullParagraph}

In fact, we shall prove Theorem~\ref{theorem:OmitAdditiveTermWFA} for a
(somewhat) larger class of \(k\)-server online algorithms, referred to as
\emph{robust} algorithms (this class will be defined soon).
We say that an online algorithm for the \(k\)-server problem is
\emph{request-sequence-oblivious}, if for every initial configuration \(A_0\),
request sequence \(\rho\), current configuration \(X\), and request \(r\), the
action of the algorithm on \(r\) after it served \(\rho\) (starting in
\(A_0\)) is fully determined by \(X\), \(r\), and the work vector
\(\WorkFunction_{\rho}(\cdot)\).
In other words, a request-sequence-oblivious online algorithm can replace the
explicit knowledge of \(A_0\) and \(\rho\) with the knowledge of
\(\WorkFunction_{\rho}(\cdot)\).
An online algorithm is said to be \emph{robust} if it is lazy,
request-sequence-oblivious, and its behavior does not change if one adds to
all entries of the work vector any given value \(d\).
We prove that if a robust algorithm is \(c\)-competitive, then it is also
strictly \((2 c)\)-competitive.
Theorem~\ref{theorem:OmitAdditiveTermWFA} follows as the work function
algorithm is robust.

In what follows, we consider a robust online algorithm \Algorithm{} and a
lazy optimal (offline) algorithm \Optimal{} for the \(k\)-server
problem.
(In some cases, \Optimal{} will be assumed to be \(X\)-lazy for some
configuration \(X\).
This will be explicitly stated.)
We also consider some underlying metric \( \Metric = (V, \Distance) \) that we
do not explicitly specify.
Suppose that \Algorithm{} is \(\alpha\)-competitive and given the initial
configuration \(A_0\), let \( \beta = \beta(A_0) \) be the additive constant
in the performance guarantee.

Subsequently, we fix some arbitrary initial configuration \(A_0\) and request
sequence \(\rho\).
We have to prove that \( \Algorithm(A_0, \rho) \leq 2 \alpha \Optimal(A_0,
\rho) \).
A key ingredient in our proof is a designated request sequence \(\sigma\)
referred to as the \emph{anchor} of \(A_0\) and \(\rho\).
Let \( \ell = \min \{ \Distance(x, y) \mid x, y \in A_0, x \neq y \} \).
Given that \( A_0 = \{ x_1, \dots, x_k \}  \), the anchor is
defined to be
\[
\sigma = (x_1 \cdots x_k)^m \text{, where }
m = \left\lceil \max\left\{ \frac{2 k \Optimal(A_0, \rho)}{\ell} + k^2,
\frac{2 \alpha \Optimal(A_0, \rho) + \beta(A_0)}{\ell} \right\} \right\rceil +
1 ~ .
\]
That is, the anchor consists of \(m\) \emph{cycles} of requests presented at
the nodes of \(A_0\) in a round-robin fashion.

Informally, we shall append \(\sigma\) to \(\rho\) in order to ensure that
both \Algorithm{} and \Optimal{} return to the initial configuration \(A_0\).
This will allow us to analyze request sequences of the form \( (\rho \sigma)^q
\) as \(q\) disjoint executions on the request sequence \( \rho \sigma \), thus
preventing any possibility to ``hide'' an additive constant in the performance
guarantee of \(\Algorithm(A_0, \rho)\).
Before we can analyze this phenomenon, we have to establish some preliminary
properties.

\begin{AvoidOverfullParagraph}
\begin{proposition} \label{proposition:RetracingCost}
For every initial configuration \(A_0\) and request sequence \(\rho\), we have
\( \Optimal(A_0, \rho, A_0) \leq 2 \cdot \Optimal(A_0, \rho) \).
\end{proposition}
\end{AvoidOverfullParagraph}
\begin{proof}
Consider an execution \(\eta\) that
(i) starts in configuration \(A_0\);
(ii) serves \(\rho\) optimally; and
(iii) moves (optimally) to configuration \(A_0\) at the end of round
\(|\rho|\).
The cost of step (iii) cannot exceed that of step (ii) as we can always
retrace the moves \(\eta\) did in step (ii) back to the initial configuration
\(A_0\).
The assertion follows since \(\eta\) is a candidate to realize \(\Optimal(A_0,
\rho, A_0)\).
\end{proof}

Since no moves are needed in order to serve the anchor \(\sigma\) from
configuration \(A_0\), it follows that
\begin{equation}
\Optimal(A_0, \rho) \leq \Optimal(A_0, \rho \sigma) \leq 2 \cdot \Optimal(A_0,
\rho) ~ .
\label{equation:TwiceTheOptimal}
\end{equation}
Proposition~\ref{proposition:RetracingCost} is also employed to establish the
following lemma.

\begin{AvoidOverfullParagraph}
\begin{lemma} \label{lemma:OfflineVisitingInitialConfiguration}
Given some configuration \(X\), consider an \(X\)-lazy execution
\(\eta\) that realizes \(\Optimal(A_0, \rho \sigma, X)\).
Then \(\eta\) must be in configuration \(A_0\) at the end of round \(t\) for
some \( |\rho| \leq t < |\rho \sigma| \).
\end{lemma}
\end{AvoidOverfullParagraph}
\begin{proof}
Assume by way of contradiction that \(\eta\)'s configuration at the end of
round \(t\) differs from \(A_0\) for every \( |\rho| \leq t < |\rho \sigma|
\).
The cost \( \Optimal(A_0, \rho \sigma, X) \) paid by \(\eta\) is at most \( 2
\cdot \Optimal(A_0, \rho) + \ConfigurationDistance(A_0, X) \) as
Proposition~\ref{proposition:RetracingCost} guarantees that this is the total
cost paid by an execution that
(i) realizes \(\Optimal(A_0, \rho, A_0)\);
(ii) stays in configuration \(A_0\) until (including) round \( |\rho \sigma|
\); and
(iii) moves (optimally) to configuration \(X\).

Let \(Y\) be the configuration of \(\eta\) at the end of round \(|\rho|\).
We can rewrite the total cost paid by \(\eta\) as \( \Optimal(A_0, \rho
\sigma, X) = \Optimal(A_0, \rho, Y) + \Optimal(Y, \sigma, X) \).
Clearly, the former term \(\Optimal(A_0, \rho, Y)\) is not smaller than
\(\ConfigurationDistance(A_0, Y)\) which lower bounds the cost paid by any
execution that starts in configuration \(A_0\) and ends in configuration \(Y\).
We will soon prove (under the assumption that \(\eta\)'s configuration at the
end of round \(t\) differs from \(A_0\) for every \( |\rho| \leq t < |\rho
\sigma| \)) that the latter term \(\Optimal(Y, \sigma, X)\) is (strictly)
greater than \( 2 \cdot\Optimal(A_0, \rho) + \ConfigurationDistance(Y, X) \).
Therefore \( \ConfigurationDistance(A_0, Y) + 2 \cdot \Optimal(A_0, \rho) +
\ConfigurationDistance(Y, X) <  \Optimal(A_0, \rho, Y) + \Optimal(Y, \sigma,
X) = \Optimal(A_0, \rho \sigma, X) \).
The inequality \( \Optimal(A_0, \rho \sigma, X) \leq 2 \cdot \Optimal(A_0,
\rho) + \ConfigurationDistance(A_0, X) \) then implies that
\( \ConfigurationDistance(A_0, X) > \ConfigurationDistance(A_0, y) +
\ConfigurationDistance(Y, X) \), in contradiction to the triangle inequality.

It remains to prove that \( \Optimal(Y, \sigma, X) > 2 \cdot \Optimal(A_0,
\rho) + \ConfigurationDistance(Y, X) \).
For that purpose, we consider the suffix \(\phi\) of \(\eta\) which corresponds
to the execution on the subsequence \(\sigma\) (\(\phi\) is an
\(X\)-lazy execution that realizes \(\Optimal(Y, \sigma, X)\)).
Clearly, \(\phi\) must shift from configuration \(Y\) to configuration \(X\),
paying cost of at least \(\ConfigurationDistance(Y, X)\).
Moreover, since \(\phi\) is \(X\)-lazy, and by the assumption that
\(\phi\) does not reside in configuration \(A_0\), it follows that in each of
the \(m\) cycles of the round-robin,
at least one server must move between two different nodes in
\(A_0\).
(To see this, recall that each server's move of the lazy
execution ends up in a node of \(A_0\).
On the other hand, all \(k\) servers never reside in configuration \(A_0\).)
Thus \(\phi\) pays a cost of at least \(\ell\) per cycle, and \( m \ell \)
altogether.
A portion of this \( m \ell \) cost can be charged on the shift from
configuration \(Y\) to configuration \(X\), but we show that the remaining
cost is strictly greater than \( 2 \cdot \Optimal(A_0, \rho) \), thus deriving
the desired inequality \( \Optimal(Y, \sigma, X) > 2 \cdot \Optimal(A_0, \rho)
+ \ConfigurationDistance(Y, X) \).

The \(k\) servers make at least \(m\) moves between two different nodes in
\(A_0\) when \(\phi\) serves the subsequence \(\sigma\), hence there exists
some server \(s\) that makes at least \( m / k \) such moves as part of
\(\phi\).
The total cost paid by all other servers in \(\phi\) is bounded from below by
their contribution to \(\ConfigurationDistance(Y, X)\).
As there are \(k\) nodes in \(A_0\), at most \(k\) out of the \( m / k \)
moves made by \(s\) arrive at a new node, i.e., a node which was not previously
reached by \(s\) in \(\phi\).
Therefore at least \( m / k - k \) moves of \(s\) cannot be charged on its
shift from \(Y\) to \(X\).
It follows that the cost paid by \(s\) in \(\phi\) is at least \( (m / k - k)
\ell \) plus the contribution of \(s\) to \(\ConfigurationDistance(Y, X)\).
The assertion now follows by the definition of \(m\), since \( (m / k - k)
\ell > 2 \cdot \Optimal(A_0, \rho) \).
\end{proof}

Since the optimal algorithm \Optimal{} is assumed to be lazy,
Lemma~\ref{lemma:OfflineVisitingInitialConfiguration} implies the following
corollary.

\begin{corollary} \label{corollary:OfflineEndingInitialConfiguration}
If the optimal algorithm \Optimal{} serves a request sequence of the form \(
\rho \sigma \tau \) (for any choice of suffix \(\tau\)) starting from the
initial configuration \(A_0\), then at the end of round \( |\rho \sigma| \) it
must be in configuration \(A_0\).
\end{corollary}

Consider an arbitrary configuration \(X\).
We want to prove that  \( \WorkFunction_{\rho \sigma}(X) \geq
\WorkFunction_{\rho \sigma}(A_0) + \ConfigurationDistance(A_0, X) \).
To this end, assume by way of contradiction that \( \WorkFunction_{\rho
\sigma}(X) < \WorkFunction_{\rho \sigma}(A_0) + \ConfigurationDistance(A_0, X)
\).
Fix \( \WorkFunction_0 = \WorkFunction_{\rho \sigma}(A_0) \).
Lemma~\ref{lemma:OfflineVisitingInitialConfiguration} guarantees that an
\(X\)-lazy execution \(\eta\) that realizes \( \WorkFunction_{\rho
\sigma}(X) = \Optimal(A_0, \rho \sigma, X) \) must be in configuration \(A_0\)
at the end of some round \( |\rho| \leq t < |\rho \sigma| \).
Let \(\WorkFunction_t\) be the cost paid by \(\eta\) up to the end of round
\(t\).
The cost paid by \(\eta\) in order to move from \(A_0\) to \(X\) is
at least \(\ConfigurationDistance(A_0, X)\), hence \( \WorkFunction_{\rho
\sigma}(X) \geq \WorkFunction_t + \ConfigurationDistance(A_0, X) \).
Therefore \(\WorkFunction_t <  \WorkFunction_0\), which derives a
contradiction, since \(\WorkFunction_0\) can be realized by an execution that
reaches \(A_0\) at the end of round \(t\) and stays in \(A_0\) until it
completes serving \(\sigma\) without paying any more cost.
As \( \WorkFunction_{\rho \sigma}(X) \leq \WorkFunction_{\rho \sigma}(A_0) +
\ConfigurationDistance(A_0, X) \), we can establish the following corollary.

\begin{corollary} \label{corollary:WorkFunctionEnding}
For every configuration \(X\), we have \( \WorkFunction_{\rho \sigma}(X) =
\WorkFunction_{\rho \sigma}(A_0) + \ConfigurationDistance(A_0, X) \).
\end{corollary}

Recall that we have fixed the initial configuration \(A_0\) and the request
sequence \(\rho\) and that \(\sigma\) is their anchor.
We now turn to analyze the request sequence \( \chi = (\rho \sigma)^{q} \),
where \(q\) is a sufficiently large integer that will be determined soon.
Corollary~\ref{corollary:OfflineEndingInitialConfiguration} guarantees that
\Optimal{} is in the initial configuration \(A_0\) at the end of round \(
|\rho \sigma| \).
By induction on \(i\), it follows that \Optimal{} is in \(A_0\) at the end of
round \( i \cdot |\rho \sigma| \) for every \( 1 \leq i \leq q \).
Therefore the total cost paid by \Optimal{} on \(\chi\) is merely
\begin{equation}
\Optimal(A_0, \chi) = q \cdot \Optimal(A_0, \rho \sigma) ~ .
\label{equation:OfflineRepetition}
\end{equation}

Suppose by way of contradiction that the online algorithm
\Algorithm{}, when invoked on the request sequence \( \rho \sigma \) from initial
configuration \(A_0\), does not end up in \(A_0\).
Since \Algorithm{} is lazy, we conclude that \Algorithm{} is not in
configuration \(A_0\) at the end of round \(t\) for any \( |\rho| \leq t <
|\rho \sigma| \).
Therefore in each cycle of the round-robin, \Algorithm{} moves at least once
between two different nodes in \(A_0\), paying cost of at least \(\ell\).
By the definition of \(m\) (the number of cycles), this sums up to \(
\Algorithm(A_0, \rho \sigma) \geq m \ell > 2 \alpha \Optimal(A_0, \rho) +
\beta(A_0) \).
By inequality~(\ref{equation:TwiceTheOptimal}), we conclude that \(
\Algorithm(A_0, \rho \sigma) > \alpha \Optimal(A_0, \rho \sigma) + \beta(A_0)
\), in contradiction to the performance guarantee of \Algorithm{}.
It follows that \Algorithm{} returns to the initial configuration \(A_0\)
after  serving the request sequence \( \rho \sigma \).

Consider some two request sequences \(\tau\) and \(\tau'\).
We say that the work vector \(\WorkFunction_{\tau}(\cdot)\) is
\emph{\(d\)-equivalent} to the work vector \(\WorkFunction_{\tau'}(\cdot)\),
where \(d\) is some real, if \( \WorkFunction_{\tau}(X) -
\WorkFunction_{\tau'}(X) = d \) for every \( X \subseteq V \), \( |X| = k \).
It is easy to verify that if \(\WorkFunction_{\tau}(\cdot)\) is
\(d\)-equivalent to \(\WorkFunction_{\tau'}(\cdot)\), then \(
\WorkFunction_{\tau r}(\cdot) \) is \(d\)-equivalent to \(
\WorkFunction_{\tau' r}(\cdot) \) for any choice of request \( r \in V \).
Corollary~\ref{corollary:WorkFunctionEnding} guarantees that the work vector
\(\WorkFunction_{\rho \sigma}(\cdot)\) is \(d\)-equivalent to the work vector
\(\WorkFunction_{\omega}(\cdot)\) for some real \(d\), where \(\omega\) stands
for the empty request sequence.
(In fact, \(d\) is exactly \( \WorkFunction_{\rho \sigma}(A_0) \).)
By induction on \(j\), we show that for every prefix \(\pi\) of \( \rho \sigma
\) and for every \( 1 \leq i < q \) such that \( |(\rho \sigma)^i \pi| = j \),
the work vector \( \WorkFunction_{(\rho \sigma)^i \pi}(\cdot) \) is
\(d\)-equivalent to the work vector \(\WorkFunction_{\pi}(\cdot)\) for some
real \(d\).
Therefore the behavior of the robust online algorithm \Algorithm{} on \(\chi\)
is merely a repetition (\(q\) times) of its behavior on \( \rho \sigma \) and
\begin{equation}
\Algorithm(A_0, \chi) = q \cdot \Algorithm(A_0, \rho \sigma) ~ .
\label{equation:OnlineRepetition}
\end{equation}

We are now ready to establish the following inequality:
\begin{align*}
\Algorithm(A_0, \rho)
& \leq ~ \Algorithm(A_0, \rho \sigma) \\
& = ~ \frac{\Algorithm(A_0, \chi)}{q} \quad \text{by
inequality~(\ref{equation:OnlineRepetition})} \\
& \leq ~ \frac{\alpha \Optimal(A_0, \chi) + \beta(A_0)}{q} \quad \text{by the
performance guarantee of \Algorithm{}} \\
& = ~ \frac{\alpha q \Optimal(A_0, \rho \sigma) + \beta(A_0)}{q} \quad \text{by
inequality~(\ref{equation:OfflineRepetition})} \\
& \leq ~ \frac{2 \alpha q \Optimal(A_0, \rho) + \beta(A_0)}{q} \quad \text{by
inequality~(\ref{equation:TwiceTheOptimal})} \\
& = ~ 2 \alpha \Optimal(A_0, \rho) + \frac{\beta(A_0)}{q} ~ .
\end{align*}
For any real \( \epsilon > 0 \), we can fix \( q = \lceil \beta(A_0) /
\epsilon \rceil + 1 \) and conclude that \( \Algorithm(A_0, \rho) < 2 \alpha
\Optimal(A_0, \rho) + \epsilon \).
Theorem~\ref{theorem:OmitAdditiveTermWFA} follows.

As the Work Function Algorithm is known to be \( (2 k - 1) \)-competitive
\cite{KP95}, we also get the following corollary.

\begin{corollary}\label{corollary:strictly}
The Work Function Algorithm is strictly \((4k-2)\)-competitive.
\end{corollary}

\paragraph{Acknowledgments} We thank Elias Koutsoupias for useful discussions. 


\end{document}